\documentclass[manuscript,acmsmall,screen]{acmart}
\graphicspath{ {img/} }
\usepackage{verbatim}
\usepackage{caption}
\usepackage{subcaption}
\usepackage{multirow}
\usepackage{rotating}
\usepackage{adjustbox}
\usepackage{balance}
\usepackage{ifthen}
\usepackage{graphicx}
\usepackage{array,booktabs}
\usepackage{tabularx}
\usepackage{textcomp}
\usepackage{makecell}
\usepackage{balance}
\usepackage{csquotes}
\usepackage{xspace}
\usepackage{verbatim}
\usepackage{varioref}
\usepackage{hyperref}
\hypersetup{
    bookmarks=false,         
    unicode=false,          
    pdftoolbar=false,        
    pdfmenubar=false,        
    pdffitwindow=false,     
    pdfstartview={FitH},    
    pdfnewwindow=true,      
    colorlinks=true,       
    linkcolor=black,          
    citecolor=black,        
    filecolor=black,         
    urlcolor=black        
}
\usepackage[capitalize,noabbrev]{cleveref}
	\Crefname{lstlisting}{Listing}{Listings}
	\crefname{lstlisting}{Listing}{Listings}
\newcommand{\refline}[2]{Line~\ref{#1:lst:#2}}
\newcommand{\reflines}[3]{Lines~\ref{#1:lst:#2} to \ref{#1:lst:#3}}

\newcolumntype{P}[1]{>{\centering\arraybackslash}p{#1}}
\usepackage{url}
\usepackage{hyperref}	
\usepackage{multicol}
\usepackage{listings}
\usepackage{color}
\usepackage[absolute]{textpos}
\usepackage[textsize=footnotesize,color=orange!20,bordercolor=gray]{todonotes}
\usepackage[binary-units]{siunitx}
\usepackage[inline]{enumitem}

\newcolumntype{Z}[1]{>{\centering\arraybackslash}m{#1}}
\newcolumntype{C}[1]{>{\centering\arraybackslash}p{#1}}
\newcolumntype{R}[1]{>{\raggedleft\arraybackslash}p{#1}}

\makeatletter
\DeclareRobustCommand{\cpp}
{\valign{\vfil\hbox{##}\vfil\cr
   \textrm{C\kern-.1em}\cr
   $\hbox{\fontsize{\sf@size}{0}\textbf{+\kern-0.05em+}}$\cr}\kern-.05em%
}
\DeclareRobustCommand{\ccpp}
{\valign{\vfil\hbox{##}\vfil\cr
   \textrm{C\kern-.1em}\cr}\raisebox{.5\depth}{/}\kern-.1em\cpp{}}
\makeatother

\colorlet{keywordcolor}{blue}
\definecolor{codebackground}{rgb}{.99,.99,.99}
\definecolor{highlightLine}{rgb}{0.85, 0.85, 0.8}
\definecolor{delim}{RGB}{20,105,176}
\definecolor{numb}{RGB}{106, 109, 32}
\definecolor{string}{rgb}{0.64,0.08,0.08}
\definecolor{classcolor}{rgb}{0.01,0.30,0.01}
\definecolor{membercolor}{rgb}{0.30,0.3,0.012}
\usepackage{calc}
\makeatletter
\newlength{\numwidth}%
\def\lst@PlaceNumber{%
  \makebox[\numwidth+.5em][l]{%
    \makebox[.75\numwidth][r]{\normalfont\lst@numberstyle{\thelstnumber}}%
  }%
}
\makeatother

\usepackage{soul}
\colorlet{codestylebg}{highlightLine!50!white}
\colorlet{codestylefg}{string}
\sethlcolor{codestylebg}
\newcommand{\codestyle}[2][codestylefg]{{\small\color{#1}\hl{\,\texttt{#2}\,}}}
\usepackage{tikz}
\newcommand*\circled[1]{\tikz[minimum height=1.1em,baseline=(char.base)]{
        \node[shape=circle,draw,outer sep=1pt,inner sep=0pt] (char) {#1};}}
\newcommand*\dotcircled[1]{\tikz[minimum height=1.25em,baseline=(char.base)]{
        \node[shape=circle,draw=red,line width=0.1em,dashed,draw,inner sep=0pt] (char) {\textbf{#1}};}}
\usepackage{scalerel}

\newcommand{\rv}{\mbox{RISC-V}\xspace}
\newcommand{\vp}{\mbox{RISC-V~VP}\xspace}

\newcommand{\ie}{i.\,e.\@\xspace}
\newcommand{\eg}{e.\,g.\@\xspace}
\newcommand{\vs}{vs.\@\xspace}
\newcommand{\cf}{cf.\@\xspace}
\newcommand{\etc}{etc.\@\xspace}
\usepackage[acronym]{glossaries}
\usepackage{glossaries-extra}

\makeatletter
\newcommand*{\glsplainhyperlink}[2]{%
    \begingroup%
    \hypersetup{hidelinks}%
    \hyperlink{#1}{#2}%
    \endgroup%
}
\let\@glslink\glsplainhyperlink
\makeatother

\setabbreviationstyle[acronym]{long-short}
\newacronym{vp}{VP}{Virtual Prototype}
\newacronym{soc}{SoC}{System-on-Chip}
\newacronym{ecu}{ECU}{Electronical Control Unit}
\newacronym{iot}{IoT}{Internet of Things}
\newacronym{cpu}{CPU}{Central Processing Unit}
\newacronym{gpu}{GPU}{Graphics Processing Unit}
\newacronym{plic}{PLIC}{Platform Level Interrupt Controller}
\newacronym{clint}{CLINT}{Core-Local Interruptor}
\newacronym{hart}{HART}{Hardware Thread}
\newacronym{eda}{EDA}{Electronic Design Automation}
\newacronym{sw}{SW}{Software}
\newacronym{hw}{HW}{Hardware}
\newacronym{hwitl}{HIL}{Hardware-in-the-Loop}
\newacronym{sdv}{SDV}{Software Driven Verification}
\newacronym{clv}{CLV}{Cross-Level Verification}
\newacronym[\glslongpluralkey={Intellectual Properties}]{ip}{IP}{Intellectual Property}
\newacronym{cisc}{CISC}{Complex Instruction Set Computer}
\newacronym{risc}{RISC}{Reduced Instruction Set Computer}
\newacronym{isa}{ISA}{Instruction Set Architecture}
\newacronym[\glslongpluralkey={Instruction Set Simulators}, \glsshortpluralkey={ISSes}]{iss}{ISS}{Instruction Set Simulator}
\newacronym{elf}{ELF}{Executable and Linking Format}
\newacronym{ic}{IC}{Integrated Circuit}
\newacronym{ide}{IDE}{Integrated Development Environment}
\newacronym{rsp}{RSP}{Remote Serial Protocol}
\newacronym{loc}{LoC}{Lines of Code}
\newacronym[plural={OSes}]{os}{OS}{Operating System}
\newacronym{dbt}{DBT}{Dynamic Binary Translation}
\newacronym{jit}{JIT}{Just-in-Time Compilation}
\newacronym{por}{POR}{Partial Order Reduction}
\newacronym{pcb}{PCB}{Printed Circuit Board}
\newacronym{rx}{RX}{Receive 'X'}
\newacronym{tx}{TX}{Transmit 'X'}
\newacronym{dift}{DIFT}{Dynamic Information Flow Tracking}
\newacronym{acm}{ACM}{Access Control Model}
\newacronym{ifp}{IFP}{Information Flow Policy}
\newacronym{lub}{LUB}{Least Upper Bound}
\newacronym{aes}{AES}{Advanced Encryption Standard}
\newacronym{pin}{PIN}{Personal Identification Number}
\newacronym{mvp}{MVP}{Minimum Viable Product}
\newacronym{usp}{USP}{Unique Selling-Point}
\newacronym{cps}{CPS}{Cyber-Physical Systems}

\newacronym{tlm}{TLM}{Transaction Level Modeling}
\newacronym{lt}{LT}{Loosely Timed}
\newacronym{at}{AT}{Approximately Timed}
\newacronym{rtl}{RTL}{Register Transfer Layer}
\newacronym{esl}{ESL}{Electronic System Level}
\newacronym{hdl}{HDL}{Hardware Description Language}
\newacronym{csr}{CSR}{Control and Status Register}
\newacronym{mmu}{MMU}{Memory Management Unit}
\newacronym{mips}{MIPS}{Million Instruction Per Second}
\newacronym{mips-processor}{MIPS}{Microprocessor without Interlocked Pipelined Stages}
\newacronym{cgf}{CGF}{Coverage Guided Fuzzing}
\newacronym{duv}{DUV}{Device under Verification}
\newacronym{dut}{DUT}{Device under Test}
\newacronym{ir}{IR}{Immediate Representation}
\newacronym{pk}{PK}{Peripheral Kernel}
\newacronym{fpga}{FPGA}{Field-Programmable Gate Array}
\newacronym{asic}{ASIC}{Application-specific Integrated Circuit}
\newacronym{tq}{TQ}{Time Quantum}

\newacronym{gui}{GUI}{Graphical User Interface}
\newacronym{api}{API}{Application Programming Interface}
\newacronym{gdb}{GDB}{GNU Project debugger}
\newacronym{gcc}{GCC}{GNU Compiler Collection}
\newacronym{tcp}{TCP}{Transmission Control Protocol}
\newacronym{udp}{UDP}{User Datagram Protocol}
\newacronym{can}{CAN}{Controller Area Network}
\newacronym{fat}{FAT}{File Allocation Table Filesystem}

\newacronym{gcov}{GCOV}{Gnu Coverage Tool}
\newacronym{ram}{RAM}{Random Access Memory}
\newacronym{dram}{DRAM}{Dynamic \acrlong{ram}}
\newacronym{spi}{SPI}{Serial Peripheral Interface}
\newacronym{mosi}{MOSI}{Master-Out-Slave-In}
\newacronym{miso}{MISO}{Master-In-Slave-Out}
\newacronym{cs}{CS}{Chip Select}
\newacronym{i2c}{I\textsuperscript{2}C}{Inter-Integrated Circuit}
\newacronym{mosfet}{MOSFET}{Metal-Oxide Semiconductor Field-Effect Transistor}
\newacronym{gpio}{GPIO}{General Purpose Input/Output}
\newacronym{led}{LED}{Light Emitting Diode}
\newacronym{oled}{OLED}{Organic \acrlong*{led}}
\newacronym{dma}{DMA}{Direct Memory Access}
\newacronym{dmi}{DMI}{Direct Memory Interface}
\newacronym{uart}{UART}{Universal Asynchronous Receiver/Transmitter}
\newacronym{pwm}{PWM}{Pulse Width Modulation}
\newacronym{adc}{ADC}{Analog-to-Digital Converter}
\newacronym{dac}{DAC}{Digital-to-Analog Converter}
\newacronym{rgb}{RGB}{Red Green Blue}
\newacronym{dsp}{DSP}{Digital Signal Processor}
\newacronym{hal}{HAL}{Hardware Abstraction Layer}
\newacronym{isp}{ISP}{In-System Programmer}
\newacronym{fifo}{FIFO}{First-In-First-Out}
\newacronym{cicd}{CI/CD}{Continuous Integration / Continuous Deployment}
\newacronym{smt}{SMT}{Satisfiability Modulo Theories}

\newacronym{ttm}{TTM}{Time-to-Market}
\newacronym{dse}{DSE}{Design Space Exploration}
\newacronym{msb}{MSB}{Most Significant Bit}
\newacronym{gcd}{GCD}{Greatest Common Divisor}
\newacronym{pnr}{PNR}{Place \& Route}
\newacronym{lc}{LC}{Logic Cells}
\newacronym{bram}{BRAM}{Block RAM}
\newacronym{tic}{TIC}{Translator Interface Controller}
\newacronym{vpil}{VPIL}{Virtual Peripheral in-the-Loop}
\makeglossaries

\title{
    Virtual-Peripheral-in-the-Loop:\\
    A Hardware-in-the-Loop Strategy to Bridge the VP/RTL Design-Gap
}

\author{Sallar Ahmadi-Pour}\authornote{These authors contributed equally to this work.}
\orcid{0000-0003-4000-6207}
\affiliation{
	\institution{Institute of Computer Science, University of Bremen}
	\streetaddress{Bibliothekstr. 5 / MZH}
	\postcode{28359}
	\city{Bremen}
	\country{Germany}
}
\email{sallar@uni-bremen.de}

\author{Pascal Pieper}\authornotemark[1]
\orcid{0000-0002-7155-2537}
\affiliation{
	\institution{Cyber-Physical Systems, DFKI GmbH}
	\streetaddress{Robert-Hooke-Str. 1}
	\postcode{28359}
	\city{Bremen}
	\country{Germany}
}
\email{Pascal.Pieper@dfki.de}

\author{Rolf Drechsler}
\orcid{0000-0002-9872-1740}
\affiliation{
	\institution{CPS DFKI GmbH}
	\streetaddress{Robert-Hooke-Str. 1}
	\postcode{28359}
}
\affiliation{
	\institution{Institute of Computer Science, University of Bremen}
	\streetaddress{Bibliothekstr. 5 / MZH}
	\postcode{28359}
	\city{Bremen}
	\country{Germany}
}
\email{drechsler@uni-bremen.de}

\copyrightyear{2023}
\acmYear{2023}

\setcopyright{none}
\acmBooktitle{ 
}
\acmPrice{
}
\acmDOI{
10.1145/xxxxxxxx.xxxxxxx
}
\acmISBN{
}

\newif\ifshady
\shadyfalse
\ifshady
\fi

\begin{document}
\ifshady
\fontsize{9}{10.23}\selectfont
\let\normalcodestyle\codestyle
\renewcommand{\codestyle}[1]{{\fontsize{9}{0}\selectfont\normalcodestyle{#1}}}
\fi


\begin{abstract}
With their ability to increase the prototyping speed and lessen the design and verification gaps, \glspl{vp} have recently gained popularity in industry and academic research.
Through \glspl{vp}, the time-to-market can be accelerated, as the system's \gls{sw} can be designed and verified before the real \gls{hw} is available.
%
Furthermore, the \gls{vp} acts as an executable specification model, offering a unified behavior reference model for \gls{sw} and \gls{hw} engineers.
%
However, between the \gls{vp} and the \gls{hw} still exists a gap, as the step from an 
architectural level \gls{vp} implementation on the \gls{tlm} to the \gls{rtl} implementation is considerably big.
%
%
Especially when a company wants to focus on their \gls{usp}, the \gls{hw} \gls{dse} and acceptance tests should start as early as possible.
Traditionally, this can only start once the (minimum) rest of the \gls{soc} is also implemented in the \gls{rtl}.
As \glspl{soc} consist of many common subsystems like processors, memories, and peripherals, 
this may impact the time-to-market considerably.
%
This is avoidable, however:
In this paper we propose a \glsxtrlong{hwitl} strategy that allows to bridge the gap between the \gls{vp} and \gls{rtl} design that empowers engineers to focus on their \gls{usp} while leveraging an existing suite of \gls{tlm} \glspl{ip} for the common base-system components.
%
We show how \glspl{vp} and partial \gls{rtl} implementations of a \gls{soc} can be combined in a \gls{hwitl} simulation environment utilizing \glspl{fpga}.
The proposed approach allows early \gls{dse}, validation, and verification of \gls{soc} subsystems, which bridges the 
\gls{tlm}/\gls{rtl} gap. 
We evaluate our approach with a lightweight implementation of the proposed protocol, and three case-studies with real-world peripherals and accelerators on \gls{hw}.
Furthermore, we assess the capabilities of our approach and offer practical considerations for engineers utilizing this \gls{hwitl} approach for \gls{soc} design; and finally propose further extensions that can boost the approach for 
specialized applications like high-performance accelerators and computation.
\end{abstract}

\maketitle


\section{Introduction}
\label{hwitl:sec:intro}
%
%
Modern \gls{soc} development is driven by the ever rising demand of a faster time-to-market for highly integrated and complex designs which are subject to phenomena known as the design gap and the verification gap~\cite{designgap,ITRS08}.
Briefly, the design gap describes the discrepancy
between an increasing capability of chip manufacturers, making more transistors per chip available, while the design engineers can not make use of that large amount of available transistors in a design~\cite{Henkel2003DesignGap}.
Similarly, the verification gap is characterized by the increasing need for verification of chip designs versus the available state-of-the-art technology for verification~\cite{Chen2017VerificationGap}.
Solving these issues 
has been the subject of research in the past decade.

Some design methodologies like \gls{esl} offer additional abstractions help to accelerate the design and verification cycles~\cite{Adamov2007ESL,Rigo2011,Rigo2011a,Santos2011}.
One of the abstractions within the \gls{esl} methodology is \gls{tlm}.
\Gls{tlm} basically abstracts away the multiple events required for a communication (like synchronizing transmitter / receiver, bit-encoding of instructions, and bus communication itself) into single transactions.
To further enhance the process of \gls{esl} based systems development a central technique is the \gls{vp}.
Unlike pure \glspl{iss}, \gls{vp}s model the structural and behavioral interaction between the processing units, the bus system and peripherals.
When \gls{vp}s are modeled with the help of the \gls{tlm} abstraction, they can gain high simulation speeds that enables booting operating systems, while analyzing 
the hardware interaction and exploring the design space efficiently.
Through \glspl{vp}, the \gls{soc} design flow is improved and allows the development of the \gls{sw} in parallel to the \gls{hw}.
As a consequence of the parallel \gls{sw} / \gls{hw} development, the time-to-market is improved and verification strategies can be employed early in the development process.
Furthermore, this allows the development team to spend more time in developing the \gls{usp} of the product.
The use of \glspl{vp} and \gls{tlm} are two key elements of the \gls{esl} methodology and are considered industry proven for the purpose of bridging the design and verification gap.
With \glspl{vp} representing executable models of the specification, \gls{hw} and \gls{sw} engineers both have well-defined interfaces with a given reference model to work with.
While the \gls{vp} will be binary compatible with the final \gls{hw} (\emph{behavioral} model), the actual \gls{hw} usually is designed with the help of \gls{rtl} abstractions.
These \gls{rtl} models can be synthesized automatically to the final layout of the transistors.
While the design and verification gap can be bridged with the \gls{esl} methodology, \glspl{vp} and \gls{tlm}, a gap between the \gls{tlm} and \gls{rtl} emerges as a visible challenge. 
Although there exist efforts in academia and industry to automatically generate \gls{rtl} code from \gls{tlm} representations, these techniques are 
mostly proprietary and not widely prevalent. 
Instead, most of the implementation is done through manual \gls{rtl} coding or with the aid of High-Level Synthesis.
On the road to a gap-free design and verification methodology, 
so called cross-level techniques have been proven to be a viable option.
Cross-level techniques that emerged in %
recent years combine %
\gls{tlm} \emph{and} \gls{rtl} representations for design and verification to leverage the advantages of the \gls{esl} methodology, specifically \gls{tlm} and \glspl{vp}.
In this paper, we propose to augment \glspl{vp} with \gls{hwitl} to add to these techniques and lessen the \gls{tlm}/\gls{rtl} gap.
Between the available \gls{vp} and the final \gls{rtl} model, we can consider parts of the system to be ready for test and verification, or even signed-off as complete. %
But without the full \gls{rtl} model available, though, these parts cannot be used and tested on the real \gls{hw}.
With the proposed methodology, the available, synthesizeable \gls{rtl} modules can be integrated to run on an \gls{fpga} while the rest of the system is still running in the \gls{vp}.
Through this technique, the \gls{tlm}/\gls{rtl} gap is bridged, as incremental results can be tested early in an integrated setup running with the real \gls{sw} applications.
This reduces the time-to-market further, as integration efforts for the final \gls{rtl} can be reduced and the focus can be shifted to the \emph{\gls{usp}} of the system (\eg a specific \gls{hw} accelerator).
As modern \gls{soc} often contain an extensive library of readily available \gls{ip}, the development of the \gls{usp} becomes the point of interest.

\begin{figure}[htp]
    \centering\includegraphics[width=\columnwidth]{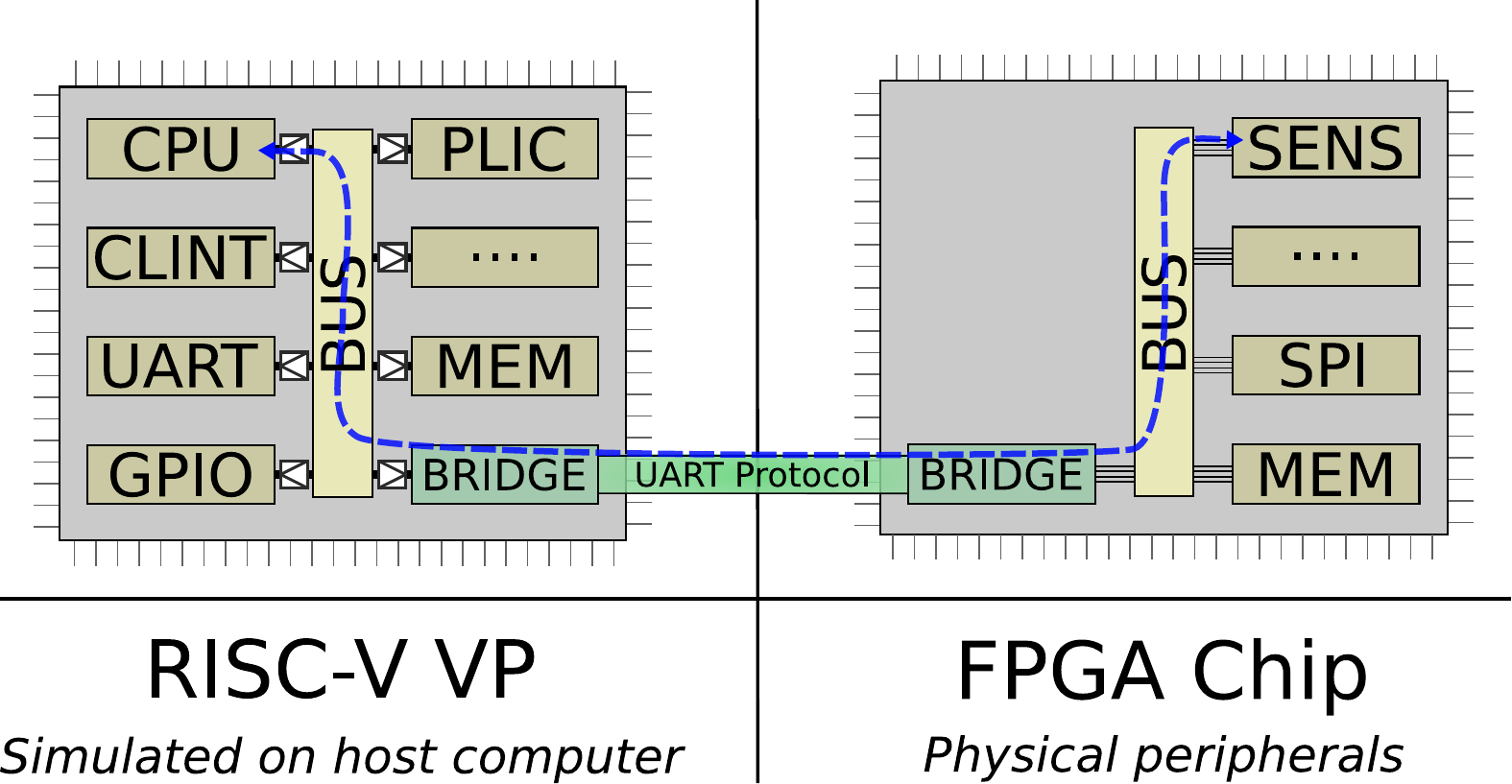}
    \caption{Architecture level overview of the proposed \glsxtrlong{vpil} approach.
		On the left side is the \gls*{tlm} virtual prototype with a memory-mapped bridge (in green) as the \textit{initiator}.
        The right side represents the real hardware with the \textit{responder} bridge handling the bus accesses.
        In blue is plotted a possible data flow path from the virtual \acrshort*{cpu} to a real sensor \gls{rtl} implementation.
        \label{hwitl:fig:arch}
    }
\end{figure}

\textbf{Contribution:}
In this paper we propose a \gls{hwitl} methodology for the \gls{soc} development flow.
It connects \gls{tlm} \glspl{vp} with memory-mapped \gls{rtl} components (specifically the \gls{usp}) while the remainder of the system's \gls{hw} is still in development.
The integration of the \gls{rtl} component is realized with a \gls{fpga} commonly used for \gls{hw} prototyping.
For the \gls{sw} running on the \gls{soc} \gls{vp}, and the \gls{hw} itself, it is fully transparent, such that integration tests can already start before the complete system is modeled in synthesizable \gls{rtl}.
A brief overview of the approach is illustrated in \cref{hwitl:fig:arch}.
A \gls{vp} model (left side) is interconnected with the \gls{fpga} based prototype (right side) through the proposed bridge (green).
The \gls{vpil} bridge allows the \gls{fpga} based prototype to work without an own \gls{cpu}, as this is part of the \gls{vp} in this example.
The proposed \gls{vpil} enables a \gls{hw}/\gls{sw} co-design boost to close the \gls{tlm}/\gls{rtl} gap with an easily extensible protocol and an \gls{ip} library of existing and well-established \gls{vp} components.
\gls{vpil} allows system designers to focus on their \gls{usp} from the design space exploration phase to early system integration tests, as well as aiding the verification with the possibility of cross-level model verification.
Additionally, it offers using the \gls{sw}-based debugging capabilities for the \gls{tlm} models, even if the synthesized chip will not have them.

In a case-study, we utilize the \vp~\cite{riscv-vp-journal} and extend it through a bridge, allowing \gls{tlm} bus transactions to be executed on a \gls{rtl} model inside an \gls{fpga} (see \cref{hwitl:fig:arch}).
In the experiments, we demonstrate how a lightweight and extensible protocol can be employed to establish a bridge between the \gls{vp} and the \gls{fpga} \gls{hw}.
We discuss considerations that arise in the implementation of the proposed bridge and what parts of the bridge can be adapted in order to fit various requirements (fast setup time, robustness, high throughput / low latency, \etc).
To stimulate further research, the proposed \gls{hwitl} bridge along with the implementation and test results of the case-study will be made available as open-source after the review process.

\section{Related Work}
\label{hwitl:sec:related}
\gls{hwitl} is a dynamic testing method that combines real \gls{hw}, simulated environments, and integrated software~\cite{istqbDefHIL}, where usually the real hardware (once available) and placed in a simulation loop where the outside environment is simulated~\cite{HIL_Overview,HIL_in_lehre}.
The strengths of this methodology enables early integration and higher quality tests as developed \gls{hw} can partially be used together with the simulation environment.
As a method, \gls{hwitl} has experienced a broad range of industrial and academic interest in the modeling of electronics~\cite{HIL_Overview,KoehlerCHILS2011,Wu2020,Jiang2004,Patrascoiu2011}, automotive~\cite{ReyesSynopsysVHIL2013,ISERMANN1999643}, aerospace~\cite{Szolc2022} and other multi-disciplinary fields~\cite{SignoreVPHIL2005,ReitzVHIL2020,LukasiewyczSysSimFPGA2014}.
In~\cite{HIL_Overview} the authors present a survey of past \gls{hwitl} approaches and the challenges in respective approaches.
The work mostly focuses on works of electrical engineering, in which control algorithms are simulated and \gls{hw} is controlled in a loop.
In \cite{Wu2020,Jiang2004,Patrascoiu2011,ISERMANN1999643,Szolc2022} \gls{hwitl} is utilized such that the \gls{sw} integrated in the final \gls{hw} is simulated (\eg the control algorithm for a plant) in the loop with prototypes of the \gls{hw}.
Such approaches are common and allow for \gls{sw} and control engineers to explore the design parameters for the control algorithm and validate and verify the control algorithms.
\cite{ReyesSynopsysVHIL2013} is an industrial approach for virtual \gls{hwitl} for automotive use, that focuses on higher level models to be integrated easily in \gls{hwitl} environments.
\cite{LukasiewyczSysSimFPGA2014} present a methodology that offers simulation and automatic optimization for distributed \gls{cps} that utilizes \gls{fpga}-based \gls{hwitl}.

\cite{KoehlerCHILS2011} describes an approach different from the mentioned ones as here the \gls{soc} is the \gls{hw} and the environment (e.g. sensor and actuator data processing are handled in a simulated environment).
\cite{ReitzVHIL2020} follows an approach similar to \cite{KoehlerCHILS2011} but addresses real-time challenges regarding data exchange and synchronization.

While none of these works directly mention a \gls{hwitl} methodology for the \gls{soc} development process, they address challenges and approaches that are common across \gls{hwitl}.
Often, the \gls{sw} part of the embedded system to simulate with \gls{hwitl} is designed either as a control system (\eg with MATLAB/Simulink), or emulated otherwise.
In these fields, a complex embedded system is commonly tightly coupled with a \gls{cps}, consisting of sensors, actuators and mechatronic systems.
The simulation of the \gls{cps} often utilizes complex simulation systems (\eg with MATLAB/Simulink), such that either the embedded system (\eg a \gls{soc}) or the physical system are modeled with \gls{hw} and respective \gls{hw} prototypes.
Through this effort, the modeling complexity within the realm of simulations can be decreased and engineers can focus on design, integration, test and verification tasks.

With technological advancements and the increase in complexity of embedded systems (specifically \glspl{soc}), industry and academia have observed the aforementioned design and verification gaps.
Naturally, engineers and researchers refine existing the design and verification processes and establish new methodologies to anticipate the gaps.
To the best of our knowledge, there exists no peripheral-centric \gls{hwitl} approach for \glspl{soc}.
One of the reasons for this may be the lack of full system hardware simulations, that only recently have been established by academia and industry.
For full system hardware simulation \glspl{vp} can be considered as an accepted methodology for complex \gls{soc} design and verification.
Due to the usage of \glspl{vp} in this work, a short overview of related works to simulators in the context of \glspl{vp} will be given 
as well.

Traditionally \gls{iss} targeting high-speed simulations such as SPIKE~\cite{riscv-spike} or QEMU~\cite{riscv_qemu} offer instruction accurate simulations.
To obtain the high performance, these simulators abstract the structure and environment modeling in their implementations.
With gem5~\cite{gem5} a full platform simulator was released, that enables inclusion of the environment interaction for the simulation.
In context of \gls{vp}-based solutions, that utilize the SystemC TLM standard for system level modeling simulators like DBT-Rise~\cite{dbt-rise} and \vp~\cite{riscv-vp-journal} were released.
Compared to aforementioned simulators TLM based simulators allow for structural system level modeling while offering flexibility to trade of between simulation speed and accuracy.
The \vp is an open-source \gls{vp} that utilizes the SystemC TLM 2.0 standard for system modeling.
Compared to DBT-Rise it offers a wide support of \rv extensions, is built in an extensible manner and has an adequate amount of documentation.

\section{Preliminaries}
\label{hwitl:sec:preliminaries}

This section will give an overview of the main systems required for the proposed \gls{hwitl} approach.
It will start with the hardware modeling framework \textit{SystemC}, followed by the rationale behind the used \gls{isa} \rv.
It finishes with an introduction of the \vp, which was chosen to host our \gls{hwitl} approach due to its open source license and widespread use.

\subsection{SystemC \gls{tlm}}
\label{hwitl:sec:sysc}%
SystemC~\cite{IEEE1666} is a hardware modeling framework that is widely adopted in the industry and academia.
It offers a \ccpp{} style modeling framework with varying degrees of timing accuracy at the benefit of simulation speed.
The structure of a SystemC design is described through ports and modules, whereas the behavior is modeled in processes which are triggered by events.
The execution of a process is non-preemptive, \ie it uses co-operative user-space scheduling for processes of each module.
This means that a process, once started, runs indefinitely until it either yields (\codestyle{wait()}) or terminates forever (return).
The process will be woken up when an event in its static sensitivity list triggers (\eg a clock edge), or it can wait for a dynamic \codestyle{sc\_event}.
This event may be triggered immediately or with a delay by, \eg, an asynchronous task, calling \codestyle{event.notify(delay)}.
For event-based synchronization, SystemC offers many variants of \codestyle{wait()} and \codestyle{notify()} such as
\codestyle{wait(time)},
\codestyle{wait(event)},
\codestyle{event.notify(delay)},
\codestyle{event.notify()},
etc.

Communication between SystemC modules can be abstracted using the \gls{tlm} standard~\cite{TLM-2.0} at the cost of timing accuracy,
but with significant improvements in simulation speed, \ie up to a factor of \num{1000} in comparison to \gls{rtl} simulation.
Especially in bus-like memory mapped communication networks, skipping interconnect procedures and signal resolutions will greatly reduce the execution time.
Instead of taking the whole route through the \gls{vp}, interactions can be initiated directly to a target port.
These transactions can either \textit{read} or \textit{write} at a specified address carrying a generic payload along with a cumulative delay, and may return either \codestyle{OK} or \codestyle{ERROR}.
This delay is increased by every model passing the transaction and added to a global quantum afterward.
The global quantum tracks the time difference a transaction \enquote{jumped} in contrast to the actual simulated time.
If this difference is bigger than the maximum allowed time, SystemC will initiate a global synchronization.
This allows for a fine control over the trade-off between simulation speed and accuracy.

\subsection{\rv}

\rv{} is an open and free \gls{isa} \cite{riscv-isa1,riscv-isa2}.
It was started in 2014 as a purely academic project and gained traction in the last years; mainly due to its extensibility and lack of licensing fees.
With the benefit of no need of backwards-compatibility, a completely new and modern \gls{isa} could be developed, reaching or outperforming existing \glspl{isa} while being more efficient in general~\cite{risc_eval}.
One of the \gls{isa}'s features is to simplify the micro-architecture by moving a part of the complexity into the compilers, as is a known \gls{risc}-approach.
This factor combined with increasing computational power in general, enables processors both in the highly specific low-power / embedded domain and the high-power server market to have a common base \gls{isa} while maintaining a clean micro-architecture.
The \rv \gls{isa} comes with a rich set of \gls{sw} and development tools, enabling developers to access state-of-the art compilers, operating systems and drivers / firmware.
Also, \rv consists of a shared base instruction set, but offers an ever-growing set of standardized and intentionally reserved (user-) instruction and \gls{csr} extensions.
This makes \rv especially suited for incorporating, \eg, neural-net and cryptography accelerators neatly and efficient into \glspl{soc} like~\cite{riscv_accelerator}.
Additionally, the ever growing ecosystem features a number of available simulators, \gls{rtl} implementations for \glspl{fpga} and \glspl{asic}, as well as full development boards for easy and quick prototyping.

\subsection{\vp}
In our work, we use and extend the implementation of an open-source \gls{vp} called \vp~\cite{riscv-vp-journal,riscv-vp-online} that is compatible to (including, but not limited to) the single-core \gls{soc} FE310\footnote{\url{https://www.sifive.com/boards/hifive1}} and the multi-core FU540\footnote{\url{https://www.sifive.com/boards/hifive-unleashed}}.
It boots Zephyr, FreeRTOS, and bare metal applications in the FE310 configuration, and Linux with peripherals in the FU540 configuration.
To improve simulation speeds the \vp offers \gls{dmi} and \gls{tq} optimizations that can be commonly found in SystemC TLM simulations.
With \gls{dmi} the performance of memory access operations is increased through using direct memory pointers instead of full TLM transactions for memory access.
Through \gls{tq} the simulator can avoid costly context switches by means of postponed synchronization with the SystemC kernel.
\section{Approach Overview}%
\label{hwitl:sec:main}

As stated earlier, the key idea of the proposed approach is to enable developing individual \gls{rtl} models without the need of designing the complete \gls{soc} first.
Following \cref{hwitl:fig:arch}, early \gls{rtl} development may already start with existing \gls{vp} \gls{tlm} models (left side).
By leveraging the proposed \gls{vpil} protocol %
with our reference implementations for the \vp and a generic \gls{fpga}, individual peripherals can directly implemented in \gls{rtl} (right side).
Bus accesses to external peripherals, initiated by the \gls{cpu}, \gls{dma}, or other bus masters on the \vp, are directly mapped via the virtual bus bridge (see \cref{hwitl:sec:PB}) and the \gls{uart} protocol (\cref{hwitl:sec:proto}).
Additionally, these forwarding memory ranges can be easily changed in the \vp via program options, allowing simultaneously existing SystemC models and on-\gls{fpga} implementations of the same \gls{ip}.
This enables a fast and easy way of behavioral cross-level testing and debugging.

The following subsections present the details of our proposed approach.
Starting in \cref{hwitl:sec:proto}, an introduction is given into the serial protocol between the \gls{vp} and a remote hardware that allows mapping these accesses to devices on the \gls{fpga}.
\cref{hwitl:sec:PB} then continues on a high abstraction %
level by explaining how virtual peripherals are connected to the \gls{vp} and how bus accesses are routed transparently.
Finally, in \cref{hwitl:sec:fpga-impl}, the implementation of the protocol decoder and bus handling on the \gls{fpga} is demonstrated.

\subsection{Protocol}\label{hwitl:sec:proto}

\begin{figure}[ht]
	\begin{subfigure}[t]{0.48\columnwidth}
	\includegraphics[width=\textwidth]{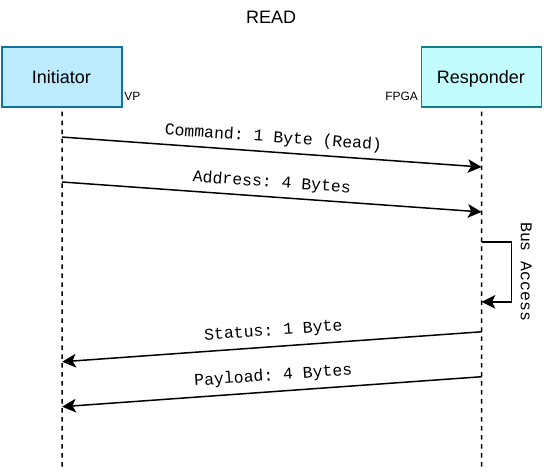}
	\caption{
		A \textit{read} request.
		The 4-Byte \textit{Payload} is always sent, including in error conditions.
	\label{hwitl:fig:read-fd}}
	\end{subfigure}
	\hfill
	\begin{subfigure}[t]{0.48\columnwidth}
    \includegraphics[width=\textwidth]{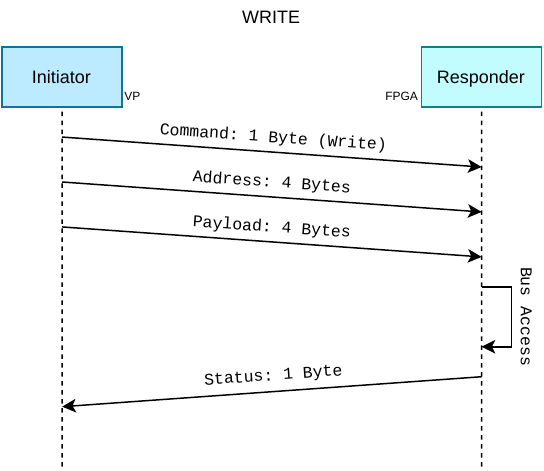}
	\caption{
		A \textit{write} request.
		\label{hwitl:fig:write-fd}}
	\end{subfigure}

	\caption{
		Flow diagrams of two requests from the \textit{initiator} to a \textit{responder}.
		All individual fields are encoded in \textit{little endian} network order.
	\label{hwitl:fig:flowdias}}
\end{figure}

\begin{lstlisting}[float=htp,
caption={
	Exerpt of the protocol data types.
	This is used by the initiator and the mock-up responder host programs.
},
label=hwitl:lst:protocol,
morekeywords= {[3],Address,Payload,Command},morekeywords = {[4]Request,ResponseStatus,ResponseRead,ResponseWrite}]
typedef uint32_t Address;
typedef uint32_t Payload;

struct __attribute__((packed)) Request {
	enum class Command : uint8_t {/*@\label{hwitl:lst:proto:command-start}@*/
		reset = 0,
		read = 1,
		write,
		getPendingIRQs,
/*@\label{hwitl:lst:proto:settime}@*/		setTime,
		exit
	} command;/*@\label{hwitl:lst:proto:command-end}@*/
	Address address;
};

struct __attribute__((packed)) ResponseStatus {
	/*
	 * Ack: bits 0 to 6
	 * irq_waiting: bit 7
	 */
	enum class Ack : uint8_t {
		never = 0,
		ok = 1,
		not_mapped,
		command_not_supported
	} ack : 7;
	bool irq_waiting : 1;
};

struct ResponseRead {
	ResponseStatus status;
	Payload payload;
};

struct ResponseWrite {
	ResponseStatus status;
};
\end{lstlisting}

The proposed \gls{vpil} protocol is a lean and hardware-parsing-friendly protocol between an \textit{initiator} (the \gls{vp}) and a \textit{responder} (the \gls{fpga} implementation).
The initiator will always start an interaction with a \codestyle[membercolor]{Command} and an \codestyle[membercolor]{Address} in network-order endianess.
Depending on the command (\eg a \textit{write}), the initiator will also transmit an \codestyle[membercolor]{Payload} (\cf \cref{hwitl:fig:write-fd}).
The responder will always respond with a \codestyle[membercolor]{Status} that contains an acknowledgment field and a flag whether an interrupt is pending or not.
In the case of a \textit{read}, it will also contain a 4 byte \codestyle[membercolor]{Payload} in network order.

Besides payload data handling, a \codestyle[membercolor]{Command} may also poll for an interrupt, reset the \gls{fpga}, and initiate other actions that are reserved for future-use (see \cref{hwitl:lst:protocol}, \reflines{hwitl}{proto:command-start}{proto:command-end}).
In case of \textit{reset} and \textit{getPendingIRQs}, no payload or address is sent in the request.
The \textit{reset} command forms a special case; as the responder immediately resets itself, no \codestyle[membercolor]{Address} is sent in the request, and no response is given.
All other commands will carry the 4-byte \codestyle[membercolor]{Address} field and expect at least one byte of \codestyle[membercolor]{ResponseStatus}.

\subsection{Peripheral Bridge}\label{hwitl:sec:PB}

The peripheral bridge is the SystemC \gls{tlm} implementation of the \textit{initiator}.
This bridge acts as a common memory-mapped bus-slave peripheral in the \vp.
Every access, however, is forwarded transparently through the communication protocol (see \cref{hwitl:sec:proto}) to a connected \textit{responder} bridge.
At this stage, the initiator bridge is unaware whether the responder is implemented in an \gls{fpga} via a serial connection, or just simulated in a separate host process.
This simplifies the testing and debugging of the protocol, and also allows extensions for future applications.

\begin{lstlisting}[float=h,
caption={
	The simplified transport function of a virtual bus member using the initiator bridge.
	Read and write accesses are mapped through the \gls{tlm}-agnostic \texorpdfstring{\codestyle[classcolor]{bus\_bridge}}{\codestyle{bus\_bridge}}.
},
label=hwitl:lst:virtual_bus_member,
morekeywords= {
	[2],hwitl,bus_bridge,not_mapped,ok
},
morekeywords= {
	[3],Address,Payload,Command,ResponseStatus,Ack
},
morekeywords = {
	[4],TLM_READ_COMMAND,TLM_WRITE_COMMAND,TLM_ADDRESS_ERROR_RESPONSE,TLM_GENERIC_ERROR_RESPONSE
}]
void VirtualBusMember::transport( tlm_generic_payload &trans, sc_core::sc_time &delay) {
	tlm_command cmd = trans.get_command();
	unsigned addr = trans.get_address();
	auto len = trans.get_data_length();

	hwitl::Payload temp = 0;/*@\label{hwitl:lst:vbm:alingedness-start}@*/
	hwitl::Payload* data = &temp;
	const bool unaligned = len != sizeof(hwitl::Payload);
	if(!unaligned) {
		data = trans.get_data_ptr();
	} else {
		if(cmd == TLM_WRITE_COMMAND)
			memcpy(data, trans.get_data_ptr(), len);
	}/*@\label{hwitl:lst:vbm:alingedness-end}@*/

	if (cmd == TLM_WRITE_COMMAND) {
		const auto response = bus_bridge.write(base_address + addr, *data);
		switch(response) {
			case hwitl::ResponseStatus::Ack::ok:
				break;
			case hwitl::ResponseStatus::Ack::not_mapped:
				trans.set_response_status(TLM_ADDRESS_ERROR_RESPONSE);
				break;
			default:
				trans.set_response_status(TLM_GENERIC_ERROR_RESPONSE);
		}
		delay += m_write_delay;
	} else if (cmd == TLM_READ_COMMAND) {
		const auto response = bus_bridge.read(base_address + addr);
		if(!response) {
			trans.set_response_status(TLM_GENERIC_ERROR_RESPONSE);
			return;
		}
		switch(response->getStatus()) {
			case hwitl::ResponseStatus::Ack::ok:
				*data = response->getPayload();
				break;
			case hwitl::ResponseStatus::Ack::not_mapped:
				trans.set_response_status(TLM_ADDRESS_ERROR_RESPONSE);
				break;
			default:
				trans.set_response_status(TLM_GENERIC_ERROR_RESPONSE);
		}
		if(unaligned)/*@\label{hwitl:lst:vbm:alingedness2-start}@*/
			memcpy(trans.get_data_ptr(), data, len);/*@\label{hwitl:lst:vbm:alingedness2-end}@*/

		delay += m_read_delay;
	}
}
\end{lstlisting}

The implementation is written in SystemC \gls{tlm}, as can be seen in \cref{hwitl:lst:virtual_bus_member}.
Read and write accesses to the peripheral are mapped through the initiator bridge, while most of the code is just for possible alignment of the four bytes of the protocol's payload (\reflines{hwitl}{vbm:alingedness-start}{vbm:alingedness-end}, and \reflines{hwitl}{vbm:alingedness2-start}{vbm:alingedness2-end}) and error handling (in the switch-cases).
The protocol handling, including the byte order packing, is completely wrapped by the virtual bus protocol implementation \codestyle[classcolor]{bus\_bridge}.
Not shown, for brevity, is the SystemC thread that periodically polls the \textit{responder} via the \codestyle[classcolor]{bus\_bridge} for pending interrupts that are then forwarded to the \gls{plic}.

\subsection{FPGA Implementation}\label{hwitl:sec:fpga-impl}

\begin{figure*}[htp]
    \centering
    \includegraphics[width=\linewidth]{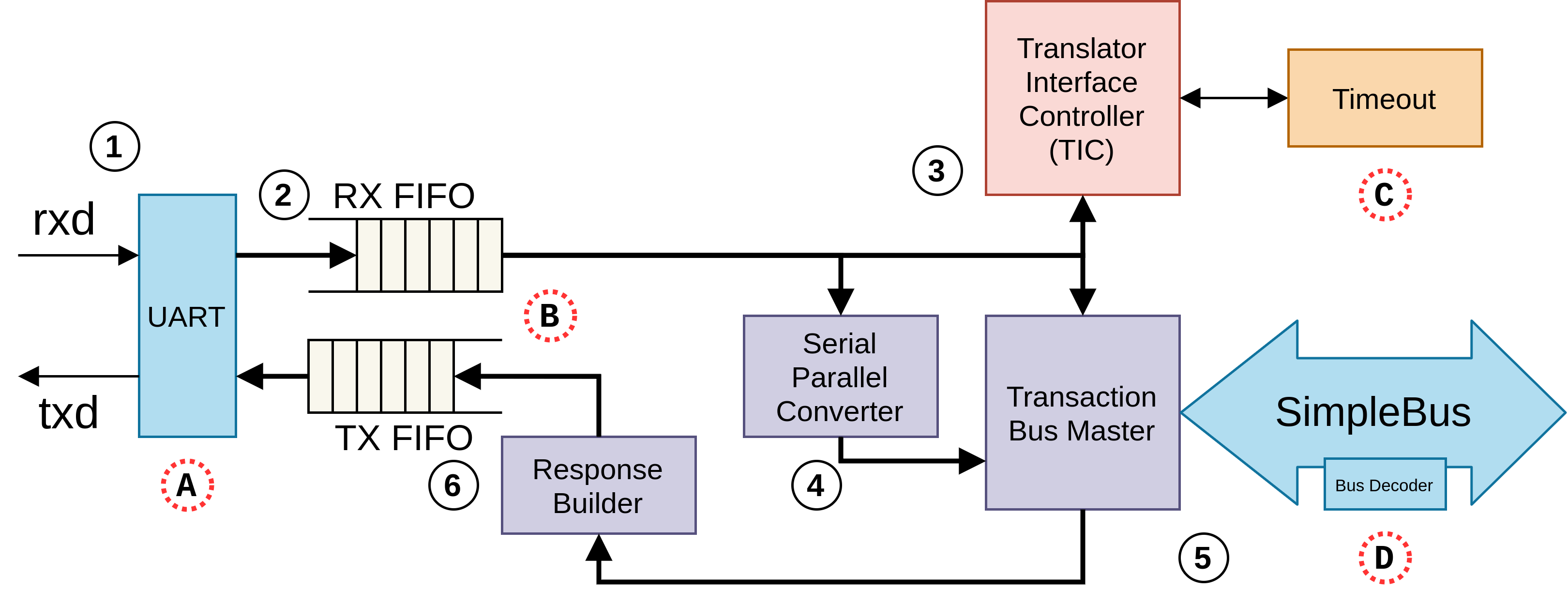}
    \caption{
        The \gls{fpga} implementation of the responder bridge.
        Modules in blue are for interfaces, models in purple represent internal modules handling communication between interfaces, and red / orange modules are for orchestration and control.
        The response buildup time is in the proposed implementation always under one millisecond.
        \label{hwitl:fig:fpga-peripheral-bridge}}
\end{figure*}
\cref{hwitl:fig:fpga-peripheral-bridge} shows a block diagram of the internal hardware architecture of the translation bridge.
The explanation will follow along the enumeration from \circled{1} to \circled{5} in \cref{hwitl:fig:fpga-peripheral-bridge} for the hardware modules and provide further information following the enumeration from \dotcircled{A} to \dotcircled{D}.
Additionally, we grouped the modules in colors to differentiate them easily for the reader.
Modules with the color blue are for interfaces, internal modules in purple are handling the bytes between interfaces, modules for orchestration and control are red and orange.
The \gls{uart} interface (\circled{1}, blue, left side) provides the byte wise serial communication with the host executing the \gls{vp}.
This interface was chosen exemplary for our case study, but any interface that will provide bytes to the internal \gls{fifo} buffer can be used.
Next the received bytes are stored in a \gls{fifo} \circled{2}, from where the bytes are directed respectively according to the protocol as discussed in \cref{hwitl:sec:proto}.
At the heart of the hardware implementation is the \gls{tic} (\circled{3}, red, top right).
The \gls{tic} orchestrates and parses the bytes according to the defined protocol, relays transactions coming from the \gls{vp} through the bus master (\circled{4}, purple, center), and handles errors like unmapped address responses and exception cases in the protocol.
Upon too much delay or other unaccounted exceptions that could stall the hardware through the interfaces, a timeout will return the system, initiated by the \gls{tic}, into a defined initial state.
As addresses and data arrive in bytes a converter (\circled{4}, purple, center), pre-processes them into chunks of 32 bit words for the transaction bus master (\circled{5}, purple, center).
The transaction bus master is attached to the \gls{soc} bus, onto which the hardware peripherals designed as \gls{rtl} modules are attached.
After processing the received transaction, the response builder (\circled{6}, purple, center) generates a protocol conform response.
The \gls{tic} handles the different cases (including errors) and can instruct the response builder to generate appropriate packets.
Generated packets are passed byte-wise into a \gls{fifo}, as the serial interface transmits the data at a different rate than the packets are generated at.
\noindent\dotcircled{A}: We used \gls{uart} for our serial interface, as it is a readily available physical layer protocol that can be extended and replaced if the requirements demand for more speed, robustness or other modes of operations.
The hardware for the serial interface can be configured for various baudrates (\eg \num{115200} baud), bit modes and additionally provide \mbox{RS-232} conforming control flow signals to allow further robustness already.

\noindent\dotcircled{B}: The \glspl{fifo} for the receiving and transmitting end are designed to be configurable for the requirements that stem from the different data rates on the serial interface and the internal processing speed.

\noindent\dotcircled{C}: For additional robustness and a configurable timeout is included.
By default, it is set to \SI{2}{\milli\second}.
If no event (such as incoming bytes, change of bus state, \etc) occurs, the timeout instructs the \gls{tic} to reset the systems state to the initial values to provide a clean and defined start.

\noindent\dotcircled{D}: The internal bus interface for our case study provides an easy to use and extensible bus configuration.
\cref{hwitl:lst:spinal-interconnect} provides an example how through the abstractions of SpinalHDL new peripherals can be easily included on the bus with a respective bus address range.
In \refline{hwitl}{spinal-interconnect:busarray} a list is declared, that will hold tuples of references to peripheral bus interfaces, the respective select signal and a bus mapping.
In \reflines{hwitl}{spinal-interconnect:peripheral-start}{spinal-interconnect:peripheral-end}, a peripheral is added by appending the list with its bus interface, select signal and respective memory mapping.
The bus mapping (\eg \codestyle{MaskMapping(0x50000000l, 0xfffffff0l)}) uses a base address and a respective mask to allow the decoder to check if a requested address maps to the peripherals base address.
This is repeated between \reflines{hwitl}{spinal-interconnect:peripheral-other-start}{spinal-interconnect:peripheral-other-end} for further peripherals and finally the list is passed in \refline{hwitl}{spinal-interconnect:decoder-start}{spinal-interconnect:decoder-end} into the bus decoder.
The bus decoder will interconnect the bus master and the list of bus interfaces according to the respective bus mapping.
This additional abstraction, made available through the features of SpinalHDL, provides an easy extension mechanism and makes the \gls{hdl} code easy to follow.

\begin{lstlisting}[float=h,
language=Scala,
caption={
SpinalHDL digest of top level peripheral bridge. Digest shows how new peripherals can be easily added to the bus infrastructure.
},
label=hwitl:lst:spinal-interconnect,
morekeywords= {
[2], ArrayBuffer, Bool, MaskMapping
},
morekeywords= {
[3],SimpleBus, GPIOLED, SBGPIOBank, SBUart, SBGCDCtrl
},
morekeywords = {
[4],val, new, ->, <>
}]
// ******** Peripherals *********
/*@\label{hwitl:lst:spinal-interconnect:busarray}@*/val busMappings = new ArrayBuffer[(SimpleBus,(Bool, MaskMapping))]

/*@\label{hwitl:lst:spinal-interconnect:peripheral-start}@*/val gpio_led = new GPIOLED() // onboard LEDs
busMappings += gpio_led.io.sb -> (gpio_led.io.sel, MaskMapping(0x50000000l,0xfffffff0l))
/*@\label{hwitl:lst:spinal-interconnect:peripheral-end}@*/io.leds := gpio_led.io.leds

/*@\label{hwitl:lst:spinal-interconnect:peripheral-other-start}@*/val gpio_bank0 = new SBGPIOBank() // GPIO for IO switches
busMappings += gpio_bank0.io.sb -> (gpio_bank0.io.sel, MaskMapping(0x50001000l,0xfffffff0l))

val gpio_bank1 = new SBGPIOBank() // GPIO for LEDs, etc.
busMappings += gpio_bank1.io.sb -> (gpio_bank1.io.sel, MaskMapping(0x50002000l,0xfffffff0l))

val uart_peripheral = new SBUart() // uart 9600 baud
busMappings += uart_peripheral.io.sb -> (uart_peripheral.io.sel, MaskMapping(0x50003000l,0xfffffff0l))
uart_peripheral.io.uart <> io.uart0

val gcd_periph = new SBGCDCtrl()
busMappings += gcd_periph.io.sb -> (gcd_periph.io.sel, MaskMapping(0x50004000l, 0xffffff00l))/*@\label{hwitl:lst:spinal-interconnect:peripheral-other-end}@*/

// ******** Master-Peripheral Bus Interconnect *********
/*@\label{hwitl:lst:spinal-interconnect:decoder-start}@*/val busDecoder = SimpleBusDecoder(
    master = busMaster.io.sb,
    decodings = busMappings.toSeq
)/*@\label{hwitl:lst:spinal-interconnect:decoder-end}@*/
\end{lstlisting}

\section{Evaluation / Case-Study}
\label{hwitl:sec:cross-level-eval}
For the case-study of the proposed approach we, envision two scenarios.
First, an incremental development in which modules are added throughout development stages, to aid the general development process for \gls{sw} and \gls{hw} developers (as real \gls{hw} descriptions become available but are not existing yet as a full system).
Second, a focused development/refinement of the \gls{usp} of a new system (\eg a specialized \gls{hw} accelerator).
State-of-the-art systems rely on a backbone of a rich and well-tested \gls{ip} library for the common, reoccurring modules.
This circumstance enables engineers to design more complex and powerful systems offering \glspl{usp} that competing chips do not offer (\eg specialized accelerators for cryptography, artificial intelligence, \etc).
Putting the focus to the \gls{usp} and making it available to the \gls{sw} developers and verification engineers, who then can design and verify the respective firmware much earlier in the design process.

\begin{figure}[htp]
    \centering\includegraphics[width=0.75\columnwidth]{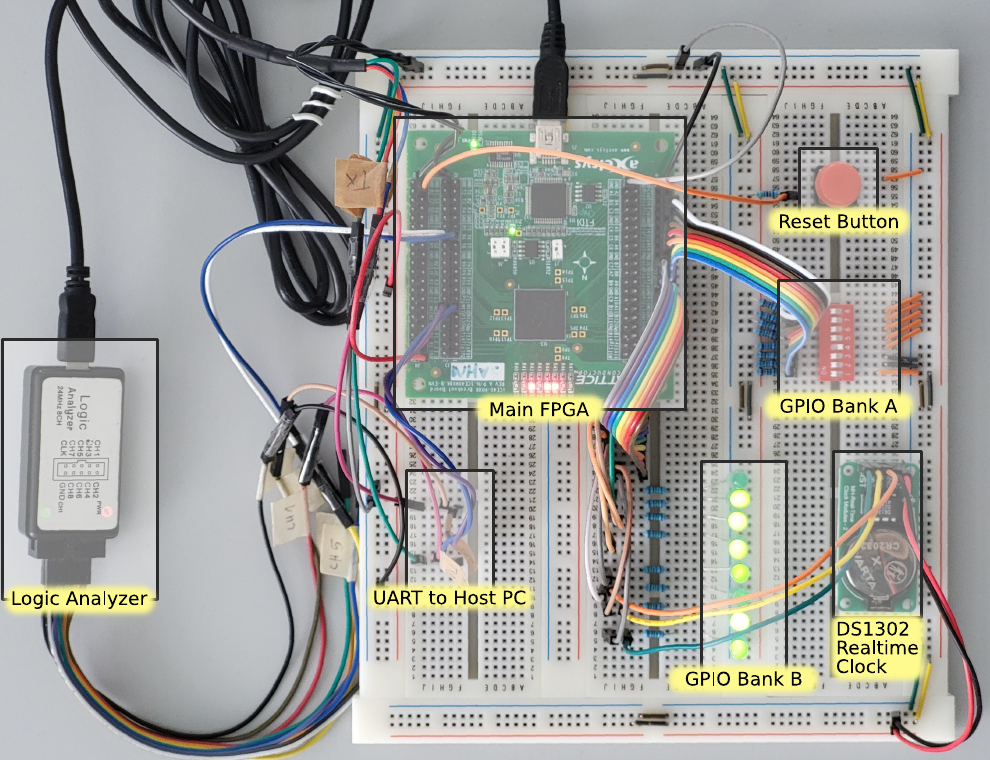}
    \caption{
        Annotated image of the experimental breadboard setup.
        The USB-connections not shown are connected to the host PC.
        \label{hwitl:fig:bb}}
\end{figure}
For the evaluation study, we use a HX8K \gls{fpga} from Lattice Semiconductor on a respective development board (see \cref{hwitl:fig:bb}).
The development board features on-board \glspl{led} (lower part of the \gls{fpga}'s \gls{pcb}) and many mappable I/O connections for prototyping.
For the experiments, these are two \gls{gpio} banks and an internal \gls{gcd} peripheral.
\Gls{gpio} bank \textbf{A} is connected to a switch array, while \gls{gpio} bank \textbf{B} is connected to \gls{led} and a DS1302 real-time clock.
The \gls{vpil} protocol is routed through the \emph{UART to Host PC} connection.
The protocol and some internal pins can be monitored through the connected logic analyzer on the left.
The used HX8K \gls{fpga} is supported through vendor tool chain as well as open source tooling.
Additionally, this model was chosen to make the approach accessible and open sourced, thus stimulating further research.
However, the choice for the HX8K \gls{fpga} also sets a limit on the available resources, so the determined area and memory usages, as well as the maximum operating frequency $f_{max}$, are more assessable.
For the host computer executing the \gls{vp}, an Intel i5-8520U @ \SI{1.60}{\giga\hertz} with Fedora 37 is utilized.

To determine the quality and effectiveness of the approach, the two scenarios will be subject to measurements of execution times, \gls{fpga} resource utilization and additional metrics.
In order to determine the time dependent behavior (\eg protocol overhead, time per transaction, \etc), for the \gls{gpio} case-study we record the transaction with the logic analyzer.
This allows measuring the duration for the read and write transactions respectively.
Additionally, the measurement allows for an estimation of the delay contributed by the \gls{fpga} environment.
Furthermore, for each case-study we measure the \gls{fpga} resource utilization in terms of area (as \gls{lc}) and memory (as \gls{bram}).
The performance and usability of the \gls{fpga} design can be estimated through the maximum operating frequency $f_{max}$ and the time it takes for the various designs to be synthesized and processed by the \gls{pnr} to obtain a bitstream for the \gls{fpga}.
For the case-study with the \gls{gcd} accelerator, the execution time of the \gls{vp} executing \gls{sw} version is compared with the \gls{hwitl} approach utilizing the \gls{rtl} implementation.

\begin{figure}[htp]
    \centering\includegraphics[height=0.5\textheight]{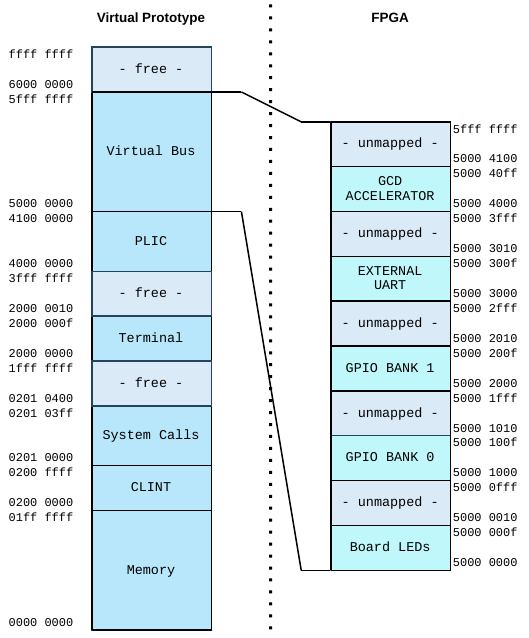}
    \caption{
        Memory map implemented for the 
        case-study.
        The simulated \gls{soc} is on the left side, while the \gls{rtl} \gls{hw} implementations are on the right side.
        \label{hwitl:fig:memory-map}}
\end{figure}

\cref{hwitl:fig:memory-map} shows the memory map implemented for the various peripherals and the exemplary accelerator.
Addresses are denoted at the sides in hexadecimal starting from the bottom (\eg Memory from \codestyle{0x0000\,0000} to \codestyle{0x01ff\,ffff}).
On the left side the memory map with its simulated peripherals inside the \gls{vp} is shown.
For the \gls{hw} implementation on the \gls{fpga} the right side shows the respective memory map.
The \gls{vpil} peripheral inside the \gls{vp} on the left side (\codestyle{0x5000\,0000} to \codestyle{0x5fff\,ffff}) is mapped transparently through the proposed protocol to the peripheral bridge on the \gls{fpga}.
In the \gls{fpga}, the memory map is implemented such that it matches the \gls{vp}'s address range.
This is not a requirement, though, as the \gls{vpil} SystemC peripheral may re-map addresses transparently.

\subsection{GPIO Bank}\label{hwitl:sec:exp:gpio}

\begin{lstlisting}[float=htp,
caption={
Simplified implementation of the \gls{gpio} bank interaction demonstration running on the \gls{vp}.
The \gls{gpio} banks are memory-mapped and behave the same as if they were implemented on the \gls{vp}.
},
label=hwitl:lst:gpio,
morekeywords= {
	[2],INTERNAL_LED,GPIO_BANK_A,GPIO_BANK_B,BUS_BRIDGE_ITR,INT_LEDs,EXT_LEDs,SWITCHES
},
morekeywords= {
	[3],BUS_BRIDGE_TYPE,MRV32_GPIO,MRV32_INTLED
},
morekeywords = {
	[4],direction,output,input,val
}]
/*@\label{hwitl:lst:gpio:memmap-start}@*/typedef uint32_t BUS_BRIDGE_TYPE;
static volatile BUS_BRIDGE_TYPE * const INTERNAL_LED = (BUS_BRIDGE_TYPE * const) 0x50000000;
static volatile BUS_BRIDGE_TYPE * const GPIO_BANK_A  = (BUS_BRIDGE_TYPE * const) 0x50001000;
/*@\label{hwitl:lst:gpio:memmap-end}@*/static volatile BUS_BRIDGE_TYPE * const GPIO_BANK_B  = (BUS_BRIDGE_TYPE * const) 0x50002000;

/*@\label{hwitl:lst:gpio:mrv32gpio-start}@*/struct MRV32_GPIO {
	volatile uint32_t direction;
	volatile uint32_t output;
	volatile uint32_t input;
/*@\label{hwitl:lst:gpio:mrv32gpio-end}@*/};
struct MRV32_INTLED {
	volatile uint32_t val;
};

static struct MRV32_INTLED* const INT_LEDs = (struct MRV32_INTLED*) INTERNAL_LED;
static struct MRV32_GPIO*   const SWITCHES = (struct MRV32_GPIO*)   GPIO_BANK_A;
/*@\label{hwitl:lst:gpio:periph-end}@*/static struct MRV32_GPIO*   const EXT_LEDs = (struct MRV32_GPIO*)   GPIO_BANK_B;

volatile static uint8_t internal_led_state = 0;
void timer_irq_handler() {
	INT_LEDs->val = internal_led_state++;
	set_next_timer_interrupt();
}

int main() {
/*@\label{hwitl:lst:gpio:somaccess}@*/	SWITCHES->direction = 0x00;
	EXT_LEDs->direction = 0xff;
	//[...]
/*@\label{hwitl:lst:gpio:endswitch}@*/	while(!(SWITCHES->input & 0b10000000)) {	// main loop
/*@\label{hwitl:lst:gpio:patternswitch}@*/		if(SWITCHES->input & 0b00000001)
			sweepLED();
		else
			countLED();
	}
	return 0;
}
\end{lstlisting}

\Cref{hwitl:lst:gpio} shows an excerpt of the basic interaction test that reads from \gls{gpio} bank \textbf{A} connected to a switch array, and writes data to the \gls{gpio} bank \textbf{B} which is connected to \glspl{led} (\cf \cref{hwitl:fig:bb}).
The global memory map is defined in \reflines{hwitl}{gpio:memmap-start}{gpio:memmap-end}, with the actual peripheral interfaces defined in \reflines{hwitl}{gpio:mrv32gpio-start}{gpio:periph-end}.
The actual read/write interaction is done through \codestyle{struct} accesses (\eg in \refline{hwitl}{gpio:somaccess}).
Based on the value of a physical, external switch (read in \refline{hwitl}{gpio:patternswitch}), the external \glspl{led} are driven in a different pattern to demonstrate the ability of interacting with the external environment.
If the switch on the \gls{msb} is unset, the program terminates (\refline{hwitl}{gpio:endswitch}).

The \gls{gpio} bank peripheral was taken from the open-source \emph{MicroRV32}~\cite{MicroRV32} that offers a set of SpinalHDL models, including a set of basic I/O peripherals.
The \gls{gpio} peripheral offers three basic 32 bit registers (see \cref{hwitl:lst:gpio}, \reflines{hwitl}{gpio:mrv32gpio-start}{gpio:mrv32gpio-end}).
The direction register determines whether a physical pin should be used for input (\codestyle{0}) or output (\codestyle{1}).
The \codestyle{input} register contains the corresponding state if input is enabled in the \codestyle{direction} register, while the \codestyle{output} register sets the physical pin state respectively.

\begin{figure}[htb]
    \centering
    \begin{subfigure}[c]{\textwidth}
        \includegraphics[width=\linewidth, trim={10px 5px 0 2px}, clip]{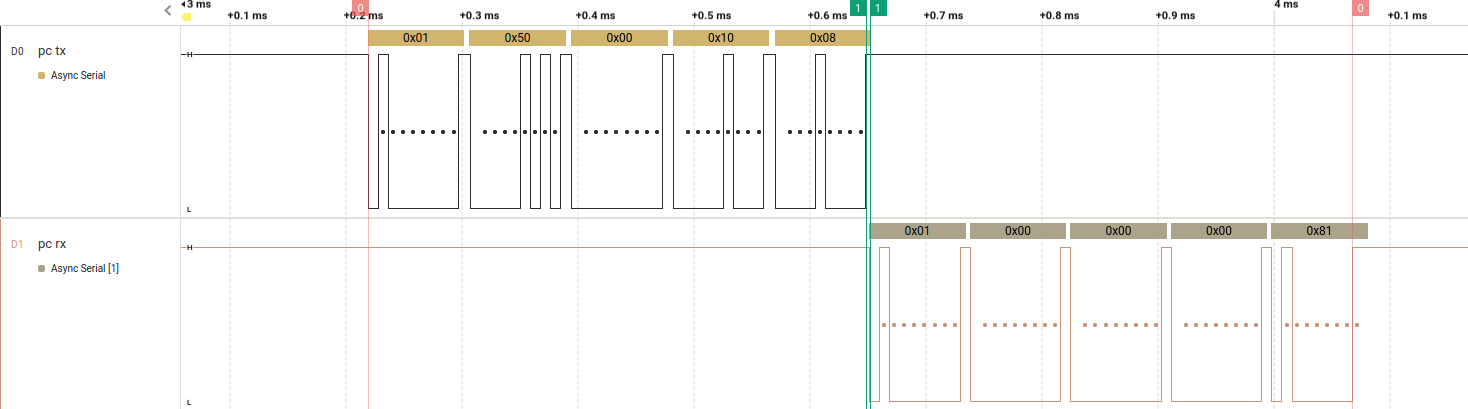}
        \caption{Read request.
            \label{hwitl:fig:readwrite:read}}
    \end{subfigure}
    \begin{subfigure}[c]{\textwidth}
        \includegraphics[width=\linewidth, trim={10px 7px 20px 5px}, clip]{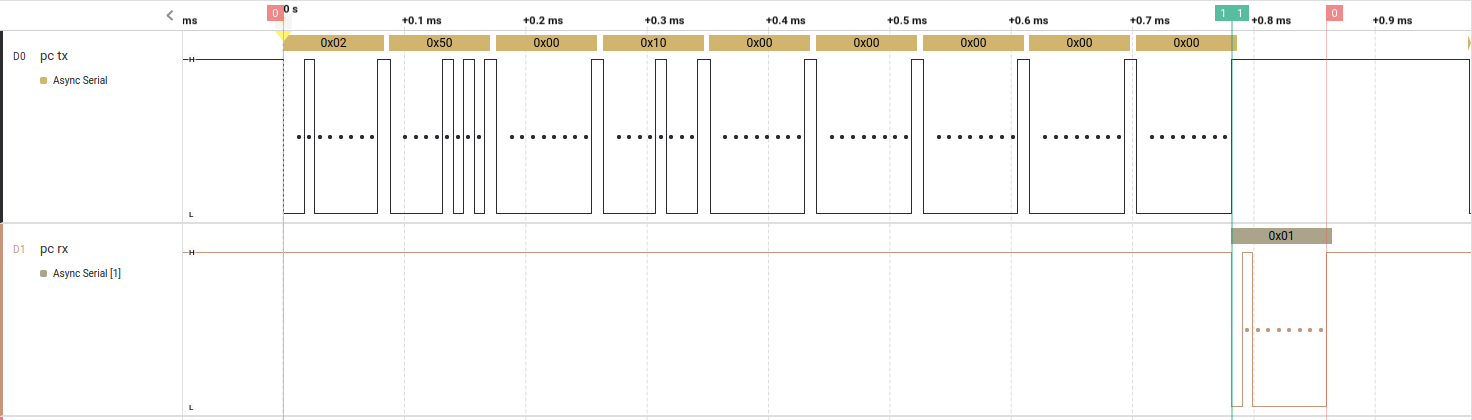}
        \caption{Write request.
            \label{hwitl:fig:readwrite:write}}
    \end{subfigure}
    \caption{
        Read (\ref{hwitl:fig:readwrite:read}) and write (\ref{hwitl:fig:readwrite:write}) transactions with annotated timing information and decoded serial communication.
        This is the \gls{uart} implementation of the proposed protocol (\cf \cref{hwitl:fig:flowdias}), and the response buildup time, in both cases, is under one millisecond (green marker \emph{1}).
        \label{hwitl:fig:readwrite}}
\end{figure}

While the functional test succeeded, the serial communication was also recorded between the \gls{vp} and the \gls{fpga} with a logic analyzer. %
For both recorded communications in \cref{hwitl:fig:readwrite}, the top portion shows the transmitted bytes from the \gls{vp} to the \gls{fpga}, while the bottom portion shows the bytes received from the \gls{fpga} as response.
The top measurement (\cref{hwitl:fig:readwrite:read}) shows a read (\codestyle{0x01}) to the address \codestyle{0x5000\,1008}, with an acknowledging response (\codestyle{0x01}) and the read data \codestyle{0x0000\,0081}.
For the whole transaction, the marker pair \emph{0} (red) indicates a time of \SI{848.25}{\micro\second}, while the internal processing on the \gls{fpga} is measured by marker pair \emph{1} (green) and takes \SI{3.25}{\micro\second}.
The bottom measurement (\cref{hwitl:fig:readwrite:write}) shows a write (\codestyle{0x02}) to the address \codestyle{0x5000\,1000} with the write data \codestyle{0x0000\,0000} and the acknowledging response (\codestyle{0x01}).
For the whole transaction the marker pair \emph{0} (red) measures a time of \SI{859.75}{\micro\second}, the internal processing on the \gls{fpga} is measured by marker pair \emph{1} (green) and takes \SI{0.626}{\micro\second}.

In this configuration, the mean protocol latency was measured as just under one millisecond at \num{115200} baud.
This is %
a promising result, as the \gls{fpga} implementation itself needs less than \SI{4}{\micro\second} and the \gls{uart} speed can be further increased if required.

\subsection{GPIO Bit-Banging SPI}\label{hwitl:sec:ds1302}

This \gls{gpio} experiment focuses on the general latency of the protocol.
In this experiment, the \gls{spi} function to interface with an DS1302 real-time clock is not implemented on the \gls{fpga} but instead bit-banged through the \codestyle{MRV32\_GPIO} bank, as introduced in \cref{hwitl:sec:exp:gpio}.
The relevant pins of the DS1302 real-time clock are \emph{CE} (chip enable), \emph{I/O} (bi-directional data port), and \emph{SCLK} (clock input for chip).
These can be used to clock-in control bytes, which are either a read- or a write command followed by an address.
With this scheme, \gls{hw}-registers can be read or written.
In the case of the DS1302, the registers contain the current time in a certain format.

For implementation, a readily available \textit{Arduino} library was used.
As it references only four functions of the Arduino framework (\codestyle{void digitalWrite(PinNumber pin, LogicLevel level)}, \codestyle{LogicLevel digitalRead(PinNumber pin)}, \codestyle{void pinMode(PinNumber pin, PinDirection dir)}, and \codestyle{void delayMicroseconds(Duration\_us duration)}), the functions could be implemented quickly to interface with the \codestyle{MRV32\_GPIO} bank.
Basically, the \gls{spi} / 3-wire protocol is implemented in \gls{sw} by setting and reading the pins, combined with accurate delays in-between.
As the \codestyle{delayMicroseconds(...)} function depends on a measure of time (through the \rv \gls{clint}), a host-time locked \gls{clint} in contrast to the usual simulation time \gls{clint} was used in this experiment.
This is needed, as the interfacing DS1302 device resides in the \enquote{real} time that needs to be synchronized.

The case-study concluded successfully as the absolute time, managed in the DS1302 chip, could be read and written over the time span of several days. 

\subsection{GCD Calculation}

To demonstrate application area for developing accelerators, a \gls{gcd} implementation in both \gls{sw} and \gls{hw} were timed against each other.
\Gls{gcd} was chosen because of the comparatively simple implementation, while still being not easy to pipeline because the length of the data-path heavily depends on the input combination.
The \gls{sw} and \gls{hw} implementation both use Euclid's algorithm to find the \gls{gcd} (see \cref{hwitl:lst:gcd}).
For the experiments, separate executables for the two implementations were build to run on the \vp.
The \gls{sw} implementation does not use the proposed \gls{vpil} bridge but implements the algorithm purely in \gls{sw} (\cref{hwitl:lst:gcd}, \reflines{hwitl}{gcd:sw-start}{gcd:sw-end}), while the \gls{hw} executable interfaces with the \gls{fpga}'s memory map (\reflines{hwitl}{gcd:hw-start}{gcd:hw-end}) tunneled through the \gls{vpil} bridge.

\begin{minipage}{.46\textwidth}
\hfill
\begin{lstlisting}[,
caption={
\Gls{sw} and memory-mapped \gls{hw} implementation of the \codestyle{gcd(a,b)} algorithm.
},
label=hwitl:lst:gcd,
morekeywords= {
	[2],GCD_ACCEL,
},
morekeywords= {
	[3],BUS_BRIDGE_TYPE,MRV32_GPIO,MRV32_INTLED
},
morekeywords = {
	[4],a,b,valid,ready,res
}]
/*@\label{hwitl:lst:gcd:sw-start}@*/uint32_t sw_GCD(uint32_t a, uint32_t b) {
	while(a != b) {
		if(a > b)
			a -= b;
		else
			b -= a;
	}
	return a;
/*@\label{hwitl:lst:gcd:sw-end}@*/}
/*@\label{hwitl:lst:gcd:hw-start}@*/uint32_t hw_GCD(uint32_t a, uint32_t b) {
	GCD_ACCEL->a = a;
	GCD_ACCEL->b = b;
	GCD_ACCEL->valid = 1;
	while(!GCD_ACCEL->ready){};
    return GCD_ACCEL->res;
/*@\label{hwitl:lst:gcd:hw-end}@*/}
\end{lstlisting}
\end{minipage}
\hfill
\begin{minipage}{.46\textwidth}
\vspace*{\fill}
\begin{table}[H]
    \centering
    \caption{
    	Test results for \gls{gcd}-implementations \codestyle{gcd(a,b)} on \gls{sw} and a memory-mapped \gls{rtl} implementation, both using Euclid's algorithm.
    	The timings include the startup- and shutdown overhead of the \vp.
	\label{hwitl:tab:gcd}}
    \begin{tabular}{rrrr}
\toprule
    	\textbf{A} & \textbf{B} & \textbf{SW [s]} & \textbf{HW [s]} \\ \midrule
    	     10154 &          3 &            0.19 &            0.17 \\
    	    101654 &          3 &            0.73 &            0.17 \\
    	   1051654 &          3 &            6.09 &            0.23 \\
    	  10512654 &          3 &           55.35 &            0.74 \\
    	     36546 &    1051654 &            0.14 &            0.17 \\ \bottomrule
    \end{tabular}
\end{table}
\hfill
\vspace*{\fill}
\end{minipage}
\Cref{hwitl:tab:gcd} shows the results of five different tests, with an increasing imbalance between the parameters \textbf{A} and \textbf{B}.
As can be seen, the \gls{sw} run-time increases faster with \emph{a}, due to the more efficient implementation on the \gls{fpga}.
The protocol overhead becomes negligible even in the sub-second execution time (for $a=\num{1051654}$ and $b=\num{3}$), although the \gls{hw} implementation uses active polling on the \gls{fpga} peripheral.

\subsection{Synthesis Results}

For the aforementioned case-studies we measured the resource utilization (area in terms of \gls{lc}, memory in terms of \gls{bram}), the maximum operating frequency $f_{max}$ and respective synthesis and \gls{pnr} times.
As the \gls{pnr} process is heuristic driven, results for the frequency and the tool run times vary for each run.
We choose to average the results over ten randomly seeded runs and provide each result with their respective standard deviation.

\begin{table*}[ht]
    \centering
    \caption{Synthesis and Place \& Route parameters for evaluated designs attached to responder bridge. Each design refers to an evaluated configuration of peripherals. Measured frequencies and times are averaged over ten runs with respective standard deviation. Area and memory utilization are shown as absolute (\#) and relative (\%) to their available resources.
        \label{hwitl:tab:synthresults}}
    {
        \begin{adjustbox}{width=\linewidth}
        \renewcommand{\arraystretch}{1.2}
        \begin{tabular}{lp{2.06cm}rrrrr}
        \toprule
        \multicolumn{2}{l}{\multirow{3}{*}[-.4em]{\textbf{Description [unit]}}}   & \multicolumn{5}{c}{\textbf{Peripheral Configuration}} \\ \cmidrule{3-7}
                                                       & &       \multirow{2}{*}{\small\textbf{\acrshort{gcd} Acc.}}       &   \multirow{2}{*}{\small\textbf{\acrshort{led}}}       &     \multirow{2}{*}{\small\textbf{\acrshort{led} + 2x\acrshort{gpio}}}	    &       \multirow{2}{*}{\shortstack[l]{\textbf{\small \acrshort{led} + 2x\acrshort{gpio}}\\\textbf{\small+ \acrshort{uart}}}}	&     \multirow{2}{*}{\shortstack[l]{\textbf{\small \acrshort{led} + 2x\acrshort{gpio}}\\\textbf{\small+ \acrshort{uart} + \acrshort{gcd}}}}
                                                        \\\\ \midrule
        \acrshort{lc} [\#]          &\multirow{2}{=}{$\Big\rbrace$\small{max.\ 7680 \acrshort{lc}}}      &       1001                    &   568                 &     706	                    &       943                             &     1432                                          \\ 
        \acrshort{lc} [\%]           &                                                &       13                      &   7                   &     9	                        &       12                              &     18                                            \\ 
        \acrshort{bram} [\#]        &\multirow{2}{=}{$\Big\rbrace$\small{max.\ 32 \gls{bram}}}      &       2                       &   2                   &     2	                        &       3                               &     3                                             \\ 
        \acrshort{bram} [\%]         &                                                &       6                       &           6           &     6                         &       9                               &     9                                             \\ 
        $f_{max}$ [MHz]               & \small {target: \SI{12}{\mega\hertz}}            &       96.86 $\pm$ 4.19        &   116.58 $\pm$ 5.62   &     113.23 $\pm$ 5.7          &       100.47 $\pm$ 3.8                &     94.97 $\pm$ 3.76                              \\ 
        \multicolumn{2}{l}{Synthesis time [s]}                                                     &       5.3 $\pm$ 0.08          &   3.93 $\pm$ 0.04     &     4.74 $\pm$ 0.08           &       6.13 $\pm$ 0.05                 &     7.48 $\pm$ 0.08                               \\ 
        \multicolumn{2}{l}{Place \& Route time [s]}                                                &       2.22 $\pm$ 0.25         &   1.24 $\pm$ 0.16     &     1.61 $\pm$ 0.22           &       2.09 $\pm$ 0.38                 &     3.41 $\pm$ 0.12                                 \\
        \bottomrule
    \end{tabular}\end{adjustbox}
    }
\end{table*}

\cref{hwitl:tab:synthresults} shows the results of the synthesis and place \& route for the utilized HX8K \gls{fpga}.
The table is split into two parts.
On the left side each description for the value is shown.
For the \gls{lc} and \gls{bram} their respective available resources on the \gls{fpga} are shown next to their description.
For the maximum operating frequency $f_{max}$, we configured the \gls{pnr} with the target frequency of \SI{12}{\mega\hertz}.
On the right side, the five columns show at first the accelerator configuration itself (second column) and then the incremental integration of additional peripherals, starting from only \glspl{led} to a configuration with four peripherals and one accelerator.
For each hardware configuration (\ie responder bridge plus respective peripherals) we collected the logic area in terms of \gls{lc} and memory \gls{bram} both in absolute and relative numbers in respect to the maximum (max.\ 7680 \gls{lc}, 32 \gls{bram}).
It should be noted, that the design with the responder bridge proves to be lightweight, as even on a small \gls{fpga} such as the HX8K the area resource utilization is small (starting with the \glspl{led} configuration at \SI{7}{\percent}).
This result emphasizes the lightweight property of the proposed \gls{hwitl} bridge.
With this, many peripherals can be attached and the integration process can be carried on for a long time into the development process to aid the engineers.
Naturally, with the incremental addition of \gls{rtl} modules, the maximum frequency $f_{max}$ decreases.

\section{Discussion}
\label{hwitl:sec:conclusion}
During development of the protocol, an appropriate focus should be given to endianess conversion.
The \ccpp{} data structures are read / written via Unix file sockets and are thus in the host endianess domain.
As the testing and validation programs for the \emph{initator} and \emph{responder} functionalities were mainly used on \texttt{x86\_64} machines (\textit{little} endian) and a certain object-oriented programming style was targeted, a part of the integration workload 
needed to be focused on synchronizing the exact byte-order between host computer and \gls{fpga}.

Further consideration should be given to the simulation \vs wall-clock time synchronization.
As the simulation may be faster or slower than the outside (or wall-clock) time, interfacing with actual devices may either require simulation-time locking (as done in \cref{hwitl:sec:ds1302}) or clock synchronization from the \gls{vp} (as is provisioned into the protocol commands, see \cref{hwitl:lst:protocol}, \refline{hwitl}{proto:settime}).
Furthermore, the case study utilized a specific physical layer (\gls{uart}) with a fixed data rate (\num{115200} baud).
The reasoning behind this was the fast setup and prototyping time, providing a proof of concept for the proposed methodology.
Switching to other protocols and techniques (\eg \gls{i2c}, Ethernet, PCIe, \etc) will drastically improve the speed, but requires additional prototyping.
Moreover, as the designs showed an already high $f_{max}$ (around \SI{100}{\mega\hertz}) on a small \gls{fpga} family (Lattice Semiconductor HX8K), an additional presumption is a boost in higher operation frequencies for bigger and faster \gls{fpga} families (\eg Xilinx Virtex, Kintex or Artix families).

These two possible enhancements can reduce the aforementioned phenomenon of synchronization, as the overhead in communication and processing can further be reduced.

Even though the case studies were implemented with \gls{uart} as the physical layer and a HX8K \gls{fpga} at \SI{12}{\mega\hertz}, the results demonstrate that the proposed methodology is a lightweight approach with adaptability for design needs towards even better speed or response times.

\section{Conclusion and Future Work}\label{hwitl:sec:fuwo}

In conclusion, this paper proposed a novel
 \gls{hwitl} strategy called \gls{vpil} that is focused on combining transaction- and register transfer layer models, effectively placing \gls{rtl} models on \glspl{fpga} \enquote{in-the-loop} of \gls{tlm} \glspl{vp}.
It leverages the existing \vp infrastructure and enables \gls{rtl} designers to focus development on their \acrlong{usp} with a minimal design evaluation cost.
The contribution includes the serial communication protocol and the respective bridge implementations in SystemC \gls{tlm} for the initiator and SpinalHDL for the \gls{fpga} responder.
The proposed approach was evaluated in separate case-studies that included modeled peripherals like \gls{gpio} banks 
and a \gls{gcd} accelerator.
To stimulate further research, the proposed tool and the case-studies will be made publicly available on GitHub in the camera-ready version (due to the blind review process).

While already proven practical, the proposed approach also opens up future work to improve the efficiency and expand the application range of \gls{vpil}:
\begin{itemize}
\item
    Use of high performance \glspl{fpga}, integrated through a \emph{PCIe} inteface, allowing for a high speed communication interface to development boards with Xilinx Virtex-7 or Artix 7 \glspl{fpga} that offer \emph{PCIe} in an \mbox{M.2} form factor.
    This would add a convenient development method on fast and high performance \glspl{fpga}, that are commonly used for artificial-intelligence accelerator development.
\item
    Utilization of \glspl{fpga} containing full \glspl{soc} (\eg Xilinx Zynq-7000 \gls{soc} series with ARM Cortex-A9), that combine configurable \gls{fpga} fabric and a commercial \gls{soc}.
    Through such \glspl{fpga}, the \gls{vp} can be executed on the accompanying \gls{soc} in a lightweight Linux environment, communicating via the \gls{vpil} protocol on the \gls{fpga} fabric.
    This could enable a flexible \gls{mvp} strategy to scale the complexity between the prototype- and small batch production phases.
\item
    Support faster interrupts besides polling by using interface mechanisms (\eg data-ready (DTR) signal from FTDI-compatible \gls{uart} devices), to improve the protocol latency for interrupts.
    If an interrupt controller is implemented on the \gls{hw} (\ie \rv's \gls{clint} or \gls{plic}), %
    this would allow a more efficient execution.
\item
	Add efficiency-improving commands to read / write at the same address again or the next higher data word.
	These would be used by buffering the target address in \gls{hw} to reduce protocol %
	overhead.
	While \codestyle{\{read,write\}\_again} would just re-use the last accessed address for improving polling the same remote register (\eg a \gls{uart} receive register), \codestyle{\{read,write\}\_consecutive} would increment the address by the register width (4 bytes) to speed up read / write accesses spanning larger address spaces (\eg filling a memory block with encrypted data).
\end{itemize}

\bibliographystyle{ACM-Reference-Format}
\bibliography{cumbib/header_long,bib/related,bib/lit-riscv,bib/hwitl_refs,cumbib/my,cumbib/related}

\end{document}